\documentclass[
twocolumn,secnumarabic,amssymb, amsmath, tightenlines,nobibnotes, aps,prl,longbibliography,superscriptaddress]
{revtex4}

\let\oldmarginpar\marginpar
\renewcommand\marginpar[1]{\-\oldmarginpar[\raggedleft\footnotesize #1]%
{\raggedright\footnotesize #1}}

\usepackage[dvipsnames]{xcolor}   
  
\usepackage[normalem]{ulem}		

\usepackage{bm}		
\usepackage{amssymb,amsmath}
\usepackage{graphicx}
\usepackage{comment}

\usepackage{enumitem}   

\usepackage{epstopdf} 
\usepackage{physics}		

\usepackage{hyperref}

\begin{document}

\title{Finite-time Landauer principle}

\author{Karel Proesmans}
\email[email: ]{Karel\_Proesmans@sfu.ca}
\affiliation{Department of Physics, Simon Fraser University, Burnaby, B.C., V5A 1S6, Canada}
\affiliation{Hasselt University, B-3590 Diepenbeek, Belgium}
\author{Jannik Ehrich}
\affiliation{Department of Physics, Simon Fraser University, Burnaby, B.C., V5A 1S6, Canada}
\author{John Bechhoefer}
\affiliation{Department of Physics, Simon Fraser University, Burnaby, B.C., V5A 1S6, Canada}

\begin{abstract}
We study the thermodynamic cost associated with the erasure of one bit of information over a finite amount of time.  We present a general framework for minimizing the average work required when full control of a system's microstates is possible.  In addition to  exact numerical results, we find simple bounds proportional to the variance of the microscopic distribution associated with the state of the bit.  In the short-time limit, we get a closed expression for the minimum average amount of work needed to erase a bit. The average work associated with the optimal protocol can be up to a factor of four smaller relative to protocols constrained to end in local equilibrium.  Assessing prior experimental and numerical results based on heuristic protocols, we find that our bounds often dissipate an order of magnitude less energy.
\end{abstract}

\maketitle

\textit{Introduction.}---Efficient computation hinges on the ability to erase a memory at minimal energetic cost. The minimum amount of work needed to complete this process is given by the \emph{Landauer limit}~\cite{landauer1961irreversibility} stating that at least $k_\mathrm{B} T \ln{2}$ of work is needed to erase a one-bit memory. Here, $k_\mathrm{B}$ is the Boltzmann constant and $T$ is the absolute temperature at which the erasure process takes place. Although Landauer's result is a cornerstone in the \emph{thermodynamics of information}~\cite{Parrondo2015_infoTD} and was the key to resolving the paradox around \emph{Maxwell's demon}~\cite{Leff19903MaxwellsDemon,Rex2017MaxwellsDemon}, it is achieved only for slow, quasi-static bit erasure.  But practical information processing requires fast erasure. 

Over the last decade, several experiments have studied the thermodynamics of slow bit erasure and have shown that one can indeed saturate the Landauer bound in the quasi-static limit \cite{berut2012experimental,jun2014high,gavrilov2016erasure,gavrilov2017direct,saira2020nonequilibrium}.  Those works, along with several theoretical studies \cite{zulkowski2014optimal,zulkowski2015optimal,boyd2018shortcuts} have suggested that the minimum amount of work needed to erase a bit over a finite amount of time is given by the Landauer limit plus a dissipative correction inversely proportional to the duration of the protocol . The associated proportionality constant, however, depends on the dynamics of the system and the constraints that one puts on the driving protocol.  Different protocols lead to different proportionality constants, raising the question of how to select the optimal protocol that minimizes this constant and hence the costs of finite-time bit erasure.

Within the fields of \emph{finite-time thermodynamics}~\cite{andresen2011current} and \emph{stochastic thermodynamics}~\cite{seifert2012stochastic,van2015ensemble} the search for protocols that minimize the average dissipation of a mesoscopic thermodynamic system during finite-time transformations has focused on the optimization of a finite (and usually small) number of control parameters influencing the potential landscape of the system~\cite{schmiedl07,bonancca2014optimal,Sivak2012TDMetrics,tafoya2019using,plata2020finite,bryant2020energy}.  For bit erasure, limiting control to a fixed set of parameters may make it more costly or even impossible to fully erase a bit \cite{diana2013finite,boyd2018thermodynamics,riechers2019balancing}.

An important advance is the work of Aurell \emph{et al.}~\cite{aurell2011optimal,aurell2012refined}, which uses full control over the potential landscape to find protocols valid in both slow and fast limits that minimize entropy production for a final state constrained to a fixed microscopic probability distribution.  However, the need to specify the final distribution is also a limitation, as the entropy production might conceivably be reduced by a different (unknown) choice of final state. 

Here, we introduce a framework that uses full control of a potential to achieve efficient, fast bit erasure without knowing in advance the ``best" final state.  Using this framework, we derive lower and upper bounds on work dissipated during optimal bit erasure.  The bounds are proportional to the initial microscopic variance of the bit and confirm that the minimum entropy production is inversely proportional to protocol duration.  We also show how to calculate the minimum work required for a given potential shape and given erasure time. Compared to previous experimental and numerical studies, taking advantage of full potential control can reduce the cost of fast erasure by roughly an order of magnitude.

In an accompanying manuscript \cite{proesmans20}, we give full details of the calculations and generalize to the case of \textit{partial erasure} of information in the bit.

\textit{Thermodynamic cost of finite-time transformations.}---Consider a bit encoded in a system described by a microscopic variable $x$.  The bit is in state $1$ if $x>0$ and state $0$ if $x<0$. The probability density $p(x,t)$ of $x$ is described by a Fokker-Planck equation,
\begin{equation}
    \pdv{p(x,t)}{t} = \pdv{x} \left(p(x,t)\pdv{x}V(x,t)\right) + \pdv[2]{x}p(x,t) \,,
\label{fpe}
\end{equation}
where $V(x,t)$ is the potential energy landscape.  In Eq.~\eqref{fpe}, we have scaled energy by $k_\text{B}T$ and lengths by $x_0 \equiv \sqrt{\textrm{Var}(x)}$, the variance of the equilibrium distribution for the potential $V(x,0) \equiv V_0(x)$. Time is scaled by $x_0^2/D$, with $D$ the diffusion coefficient. This description applies to a broad class of systems, including colloidal particles trapped in a potential \cite{berut2012experimental,jun2014high,gavrilov2016erasure,gavrilov2017direct} and superconducting fluxes \cite{saira2020nonequilibrium}.  In such systems, the microscopic state $x(t)$ is coarse-grained to two (or more) macrostates that encode information \cite{landauer1961irreversibility}.

Building on ideas from stochastic thermodynamics \cite{seifert2012stochastic,van2015ensemble} and optimal-transport theory \cite{villani2003topics,levy2018notions}, one can calculate the minimum average work to go from an initial microscopic equilibrium distribution $p_0(x)$ to a final microscopic distribution $p_\tau(x)$ over a time interval of length $\tau$ for protocols having the same start and end point, with $V(x,0)=V(x,\tau) \equiv V_0(x)$.  Assuming full control over the potential $V(x,t)$, one finds \cite{aurell2011optimal,aurell2012refined,zhang2019work,zhang2020optimization} 
\begin{multline}
    W = \underbrace{\int_{-\infty}^{\infty}\dd{x}\, 
    p_\tau(x)\ln\frac{p_\tau(x)}{p_0(x)}}_{\Delta \mathcal{F}} \\
    + \underbrace{\frac{1}{\tau} \int_0^1 \dd{y} \,         \left[ f^{-1}_0(y)-f^{-1}_\tau(y)
        \right]^2}_{\Delta_\textrm{i}S} \,,
\label{wminmicro}
\end{multline}
where $f_{0/\tau}(x)=\int_{-\infty}^x \dd{x'} \, p_{0/\tau}(x')$ are the associated cumulative distributions and $f^{-1}$ their inverses.  The quantity $\Delta \mathcal{F}$ is the change in nonequilibrium free energy arising solely because the probability density is transformed from $p_0(x)$ to $p_\tau(x)$, and $\Delta_\textrm{i}S$ is the average entropy production of the transformation~\cite{Esposito2011NonequilibriumFreeEnergy}. A similar expression holds for discrete systems \cite{shiraishi2018speed,zhang2018comment}

For many practical applications, one is interested not in the exact ``microscopic" distribution of the system but rather in a coarse-grained, ``macroscopic" distribution. For example, when erasing a bit of memory, one is not interested in the full distribution of the microscopic variable but only in the probability for the bit to be in macrostate $0$, $P_\textrm{L}=p(x<0)$ or in macrostate $1$, $P_\textrm{R}=p(x>0)$.  This means that fully minimizing the amount of work to go from an initial macroscopic distribution to a different final macroscopic distribution implies a \textit{second} minimization of Eq.~\eqref{wminmicro} over all possible microscopic distributions $p_\tau(x)$ that are compatible with the desired final macroscopic distribution.  (The initial distribution $p_0(x)$ is fixed if we start in thermal equilibrium.)   Following \cite{aurell2011optimal,aurell2012refined,zhang2019work,zhang2020optimization}, we change variables from $f_\tau(x)$ to $\Gamma(x) \equiv f_\tau^{-1}\left(f_0(x)\right)$.  The minimum amount of work required to complete the process is then
\begin{multline}
    W_\textrm{min} = \min_{\Gamma(x)}
    \int_{-\infty}^{\infty} \dd{x} \, p_0(x) \, \times \\
    \left[\ln \frac{p_0(x)}{\Gamma'(x)p_0(\Gamma(x))}
    +\frac{\left[\Gamma(x)-x\right]^2}{\tau} \right] \,,
\label{gamw}
\end{multline}
where the minimization is done over all $\Gamma(x)$ that correspond to the correct macroscopic final distribution. For ``full" bit erasure at time $\tau$, the particle is always somewhere within the region corresponding to macrostate 0 ($x<0$), so that $P_\textrm{L}=1$ and $P_\textrm{R}=0$ for $t=\tau$. Consequently,
\begin{equation}
    f_\tau(0)=1, \quad \textrm{or} \quad \Gamma\left(f_0^{-1}(1)\right)=0 \,,
\end{equation}
where the first condition again implies that all probability density is in $x<0$ and the second implies that at $t=0$, we have $f_0(\infty)=1$ and hence $\Gamma(\infty)=0$, since $\Gamma$ maps positions at time $\tau$ to positions at $t=0$.  For boundary conditions appropriate to partial erasure, see Ref.~\cite{proesmans20}.

Using the calculus of variations, we find that the optimal $\Gamma(x)$, and therefore the optimal final microscopic distribution, obeys the Euler-Lagrange equation \cite{proesmans20}
\begin{equation}
    V'(\Gamma(x)) - \frac{V'(x)}{\Gamma'(x)} - \frac{\Gamma''(x)}{\Gamma'(x)^2} + \frac{2}{\tau}\left[ \Gamma(x)-x \right] = 0 \,,
\label{cons}
\end{equation}
where  we have assumed that at $t=0$, the system is in equilibrium, $p_0(x)\sim\exp\left(-V_0(x)\right)$.

\textit{Bounds on finite-time bit erasure.}---Having formulated a general theory, we apply it to the problem of bit erasure. We consider a system described by Eq.~\eqref{fpe}, with a potential that is initially symmetric, $V_0(x)=V_0(-x)$.   Although solving the minimization condition, Eq.~\eqref{cons}, cannot in general be done analytically, we can nonetheless place upper and lower bounds on $W_\textrm{min}$. 

\begin{figure}[ht]
    \includegraphics[width=3.2in]{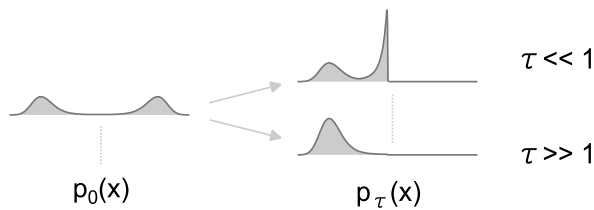}
    \caption{Optimal protocols in the short- and long-time limits.  Starting from the equilibrium distribution $p_0(x)$ for a symmetric potential $V_0(x)$ at time $t=0$ (left distribution), the system state is transformed to $p_\tau(x)$ at time $\tau$.  For $\tau \ll 1$ (upper right distribution), the final distribution is approximately the sum of the initial probability density of the left well plus a sharp peak composed of probability transported from the right well.  For $\tau \gg 1$ (lower right distribution), the final distribution is in local equilibrium in the left well.}
\label{fig:twoLimits}
\end{figure}

To establish an upper bound for $W_\textrm{min}$, we fix the final microscopic distribution to be the local-equilibrium distribution that fixes all probability to be in the region $x<0$ (Fig.~\ref{fig:twoLimits}, bottom right),
\begin{equation}
    p_{\tau}(x)=p_{\textrm{leq}}(x) =
    \begin{cases}
        2p_0(x) \,, & x<0 \,,   \\
        0       \,, & x>0 \,.
    \end{cases}
\end{equation}
The local equilibrium distribution minimizes the first term in Eq.~\eqref{gamw}, in accordance with the boundary condition of full erasure.

The optimal protocol for this case leads to an average work $W_{\textrm{min,leq}}$. We have $W_{\textrm{min}}\leq W_{\textrm{min,leq}}$, because constraining the final distribution to a local equilibrium can only increase the work relative to the case where we allow the final distribution to be selected from a set of distributions with $p_\tau(x > 0) \equiv 0$. In Ref.~\cite{proesmans20}, we show that
\begin{equation}
       W_\textrm{min,leq} \leq \ln 2 + \frac{2}{\tau} \,.
\label{upbound}
\end{equation}
In Ref.~\cite{proesmans20}, we also derive an alternate lower bound for $W_\textrm{min,leq}$ based on Ref.~\cite{dechant2019thermodynamic}.

To derive a lower bound on $W_{\textrm{min}}$, we observe that Eq.~\eqref{gamw} minimizes the sum of two terms.  Minimizing each term separately then gives a lower bound:
\begin{widetext}
\begin{align}
    W_\textrm{min} &\geq \min_{\Gamma(x)}
    \int_{-\infty}^{\infty} \dd{x} \, p_0(x) \,
   \ln \frac{p_0(x)}{\Gamma'(x)p_0(\Gamma(x))} 
   + \min_{\Gamma(x)}\int_{-\infty}^{\infty} \dd{x} \, p_0(x) \frac{\left[\Gamma(x)-x\right]^2}{\tau} \nonumber \\[3pt]
   &= \min_{p_\tau(x)}\int_{-\infty}^{\infty} \dd{x} \, 
        p_\tau(x)\ln\frac{p_\tau(x)}{p_0(x)} 
        + \frac{1}{\tau} \int_0^\infty \dd{x} \, p_0(x)x^2  \nonumber \\
   &= \ln 2 + \frac{1}{2\tau} \,.
\label{lowbound}
\end{align}
\end{widetext}

As with the upper bound, the first term of Eq.~\eqref{lowbound} is also minimized by the local-equilibrium distribution.  By contrast, the optimal choice of $\Gamma$ in the second term is $\Gamma(x)=x$ for $x<0$ (which minimizes the integrand) and $=0$ for $x\ge0$ (because no probability is left at the end for $x\ge0$).  More visually, the optimal protocol for the second term ``pushes" the probability initially in the right well to a spike at $x \approx 0^{-}$.  The probability in the ``wrong" well is moved the minimum amount possible to be in the correct macrostate---pushed to its edge---while the probability already in the macrostate is left untouched (Fig.~\ref{fig:twoLimits}, top right).  As $\tau \to 0$, the spike of probability at the edge of the macrostate approaches a delta function. 

Piling the probability into a delta function leads to an infinite contribution from the first term, since the free energy of a perfectly localized particle is infinite; however, in the limit $\tau \to 0$, the second term has an infinite weight $\tau^{-1} \to \infty$, allowing for such singular behavior.

As the main result of this paper, we rewrite these bounds in terms of the original unscaled quantities:
\begin{equation}
    \ln 2+\frac{\textrm{Var}(x)}{2D\tau} \leq
        \frac{W_\textrm{min}}{k_\textrm{B}T}
        \leq \frac{W_\textrm{min,leq}}{k_\textrm{B}T}
        \leq \ln2+\frac{2\textrm{Var}(x)}{D\tau} \,.
\label{febounds}
\end{equation}
That is, the cost to fully erase a bit over a finite amount of time is equal to the Landauer cost, $\ln 2$, plus a term that is determined by the initial variance of the distribution.  Remarkably, for all $\tau$, the minimum entropy production is always $\sim \tau^{-1}$.  We notice, in particular, that the upper and lower bounds to the entropy production differ by a factor of four.  We can understand this numerical factor by noting that an approximate expression for the dissipation is
\begin{align}
   \frac{F\Delta x}{k_\textrm{B}T}
        \sim \frac{\gamma}{k_\textrm{B}T} 
            \left( \frac{\Delta x}{\tau} \right)\Delta x 
        \sim \frac{(\Delta x)^2}{D\tau} \,,
\end{align}
where we apply the friction force $F = \gamma \dot{x} \approx \gamma (\Delta x/\tau)$ and the Einstein relation, $D=k_\textrm{B}T/\gamma$.  The quantity $\Delta x$ is the typical distance a particle is transported during the protocol.  In the long-time limit, the system stays in local equilibrium, and the probability from the right well is shifted to the left well.  In the short-time limit, the same probability is moved only half as far (by symmetry of the potential) to $x=0$.  The factor-two reduction in $\Delta x$ decreases the dissipation by a factor of four.

In special cases, we can saturate the bounds in Eq.~\eqref{febounds}.  In the fast-erasure limit $\tau \ll 1$, the $\tau^{-1}$ term in Eq.~\eqref{lowbound} dominates, leading to saturation at the lower bound in Eq.~\eqref{febounds}, giving a general, closed expression for the cost of fast erasure of a bit.  In the slow-erasure limit $\tau \gg 1$, the erasure cost reduces to the Landauer cost, as expected.  For a first-order correction in $1/\tau$, one can verify that $W_{\textrm{min}}=W_{\textrm{min,leq}}$, saturating the second inequality in Eq.~\eqref{febounds} \cite{proesmans20}. Finally, the last inequality in Eq.~\eqref{febounds} is saturated for an initial ``two-state" distribution,
\begin{equation}
    p_0(x) = \frac{1}{2} 
        \left[ \delta \left( x-\tfrac{1}{2}\Delta x \right) 
        + \delta \left( x+\tfrac{1}{2}\Delta x \right) \right] \,,
\label{twostate}
\end{equation}
where $\Delta x$ is the difference in $x$ between the two states. Equation~\eqref{twostate} is the limiting distribution for a broad class of double-well potentials with an infinite barrier between the wells.

\begin{figure}[ht]
    \includegraphics[width=2.6in]{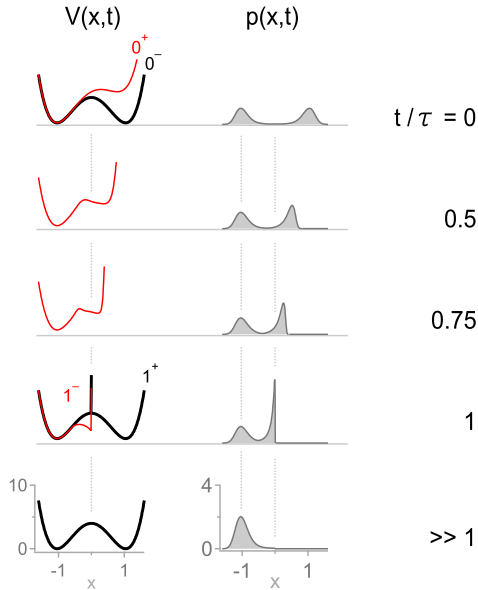}
    \caption{Erasure protocol for $\tau=0.2$ and $E_\textrm{b}=4$.  The original and final double-well potentials $V_0(x)$ are shown in black at the top and bottom of the left column.  Red potentials $V(x,t)$ denote the control.  The control is carried out for $0^+ < t/\tau < 1^-$, and the potential changes discontinuously at $t=0$ and $t/\tau=1$.  At $t/\tau=1$, an infinitesimally narrow, extra barrier $\delta(0)$ is added to $V_0(x)$ to prevent probability from leaking back into the right well.  It is removed at a later time $t/\tau \gg 1$, after the system has relaxed to local equilibrium.  The right-hand column shows corresponding probability distributions.}
\label{bitErase_col02}
\end{figure}

\textit{Example.}---Let the initial energy landscape be (Fig.~\ref{bitErase_col02})
\begin{equation}
    V_0(x) = E_\textrm{b}
        \left[ \left( \frac{x}{x_\textrm{m}} \right)^2 - 1 \right]^2 \,,
\label{Vdw0}
\end{equation}
with a barrier $E_\textrm{b}=4$ between the wells and $x_\textrm{m} \approx 1.04$, which implies an  equilibrium variance Var$(x)=1$.  Figure~\ref{bitErase_col02} shows the optimal protocol $V(x,t)$ and corresponding densities $p(x,t)$ for $\tau=0.2$.  The protocol has jump discontinuities when passing from $V_0(x)$ (black curve) at $t=0^-$ to $V(x,t=0^+)$ (red curve) and similarly in passing from $V(x,t=1^-)$ to $V_1(x)=V_0(x)$.  At $t=\tau$, we add a $\delta(0)$ barrier that keeps probability from leaking back into the right well for $t>\tau$.  Notice that no work is done for $t>\tau$.  The probability that is trapped in the left well then relaxes to a local equilibrium, after which the barrier may be removed. See the bottom plots.  

\begin{figure}[ht]
    \includegraphics[width=3.0in]{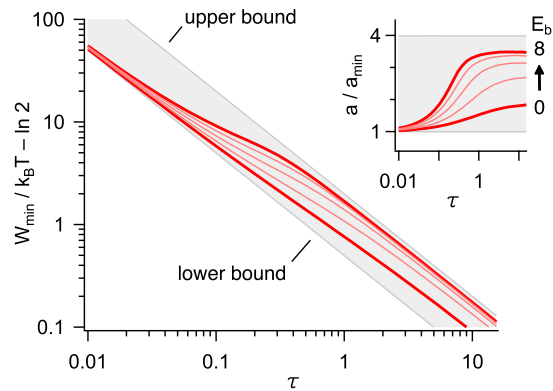}
    \caption{Minimum entropy production in excess of Landauer bound.  The shaded regions show upper ($2\tau^{-1}$) and lower ($\tau^{-1}/2)$ bounds. The inset shows  $a \equiv \tau(W_\textrm{min}/k_\textrm{B}T-\ln 2)$ relative to its lower bound.  Red curves are plotted for $E_\textrm{b} = \{0,2,4,6,8\}$.  Heavier lines denote the $E_\textrm{b}=0,8$ cases.}
\label{fig:workEx}
\end{figure}

We then numerically calculate the upper and lower bounds, Eqs.~\eqref{upbound} and \eqref{lowbound}, and $W_{\textrm{min}}$ and $W_{\textrm{min,leq}}$ for full erasure. Figure~\ref{fig:workEx} shows that the upper and lower bounds are satisfied.  We also note that  $W_{\textrm{min}}\approx W_{\textrm{min,leq}}$ in the slow-driving limit and that $W_{\textrm{min}}$ saturates the lower bound in the fast-driving limit. For $E_\textrm{b} \gg 1$, the potential wells are quite steep, and the Boltzmann distribution resembles quite well the two-delta-function distribution, Eq.~\eqref{twostate}, which explains why $W_{\textrm{min,leq}}$ is close to the upper bound.

\textit{Comparison with experimental and numerical results.}--- Over the last decade, several high-precision tests of the Landauer principle have been performed \cite{berut2012experimental,berut2013detailed,berut2015information,jun2014high,gavrilov2017direct}. In general, those protocols satisfied $W/k_\textrm{B}T-\ln 2\sim \tau^{-1}$ in the slow-driving limit. Therefore, for large $\tau$, the measured work in those experimental protocols has qualitatively the same form as the optimal protocol, raising the question of how close the experimental results are to the optimum. From Table \ref{tab:literature}, we can see that the measured amount of entropy production exceeds the optimum by factors of 2--6 (see Supplemental Material \cite{Proesmans20SI}).

Furthermore, we can also compare our bound to numerical studies of bit erasure. Zulkowski and DeWeese \cite{zulkowski2014optimal} calculate the amount of work to erase a bit in a potential consisting of two flat wells of length $\ell$, separated by a thin wall of arbitrary height. If one  controls only the height of the wells and uses slow driving, they showed that the minimum amount of work to erase a bit is given by $D\tau/{\textrm{Var}(x)}({W}/{k_\textrm{B}T}-\ln 2)=3\left((\sqrt{2}-1)^2+1\right)\approx 3.51$. By contrast, Boyd \emph{et al}.~\cite{boyd2018shortcuts} derived a general framework to calculate the work to erase a bit via a protocol that keeps the system always in local equilibrium. For the flat-well potential used by Zulkowski and DeWeese, this protocol actually performs better than the limited-control protocol used in \cite{zulkowski2014optimal}, $D\tau/{\textrm{Var}(x)}({W}/{k_\textrm{B}T}-\ln 2)=\pi^2(2-\sqrt{2})/2 \approx 2.89$. However, for a double-well potential of the form of Eq.~\eqref{Vdw0} and $E_\textrm{b}=10$, applying the method of Ref.~\cite{boyd2018shortcuts} leads to average work values that are several orders of magnitude larger.

\begin{table}[htb]
\caption{\label{tab:literature}%
Comparison between work measurements found in the literature and $W_\textrm{min}$ in terms of ${\textrm{Var}(x)}/{D\tau}$ \cite{Proesmans20SI}.}
\begin{ruledtabular}
\begin{tabular}{@{\hspace{1em}} l c c c @{\hspace{1em}}}
Experiment/Numerics &$\frac{W}{k_\textrm{B}T}-\ln 2$ &  $\frac{W_\textrm{min}}{k_\textrm{B}T}-\ln 2$ & ratio\\
\colrule
Bérut \emph{et al}.~\cite{berut2012experimental,berut2013detailed,berut2015information} &10.2 & 1.80 & 5.67\\
Gavrilov \emph{et al}.~\cite{gavrilov2017direct} & 7.20 & 1.82 & 3.96\\
Jun \emph{et al}.~\cite{jun2014high} & 5.67 & 1.82 & 3.11\\
Zulkowski \emph{et al}.~\cite{zulkowski2014optimal}& $3.51$ & 1 & 3.51\\
Boyd \emph{et al}.~\cite{boyd2018shortcuts} &$ 2.89 $& 1 & 2.89
\end{tabular}
\end{ruledtabular}
\end{table}

All the above protocols were explored in the slow ($\tau \to \infty$) limit. But we have shown here that the entropy production for optimal protocols, when scaled by ${\textrm{Var}(x)}/{D\tau}$, drops to $1/2$ for fast driving  ($\tau\rightarrow 0$). Thus, for fast erasure, our protocol can improve efficiency by up to a further factor of four.

\textit{Conclusions and outlook}.---When erasing a bit, dissipation is minimized by moving probability as little as possible, given the final-state constraint.  Long protocols are automatically in local equilibrium, but short protocols can increase performance by moving probability to the edge of the desired macrostate.  In one dimension, the move is half the distance compared to one that maintains local equilibrium, reducing dissipation by up to a factor of four.

We suggest three extensions of our formalism: 
\begin{itemize}[leftmargin=*]
\setlength\itemsep{-0.1em}

\item \textit{Higher dimensions}.  The factor-four improvement results from the one-dimensional geometry.  

\item \textit{Reduced damping}.  Bit erasure might be more efficient for critically damped systems \cite{deshpande2017designing}.  
\item \textit{Quantum effects}. Landauer's principle also holds for quantum systems strongly interacting with their environment \cite{reeb2014improved,mohammady2016minimising}.

\end{itemize}

A remarkable feature of the optimal solutions is the existence of various discontinuities and singularities in the control.  Here, as elsewhere \cite{schmiedl07}, there are discontinuities in the potential at the beginning and end of the protocol; in addition, the intermediate-time potential can have near-discontinuities in the slope \cite{aurell2012refined}, which become more pronounced for fast driving (Fig.~\ref{bitErase_col02}).  Unfortunately, experimental systems are likely unable to reproduce such protocols exactly \cite{martinez2016engineered,chupeau2018thermal}.  Moreover, optimal protocols assume a ``perfect" model of the system under control.  But parameters are always uncertain, and the shape of the underlying potential $V_0(x)$ may simplify a more complex reality.  For such reasons, experiments can only approximate the optimal solutions derived here.  The challenge---and this is what makes optimal control a problem of physics as well as mathematics---is to find good approximations to the ``best" control \cite{gingrich2016near} that are robust to imperfections of system models and to experimental constraints.

We thank David Sivak and Raph\"el Ch\'etrite for helpful comments.  
This work was supported by a Foundational Questions Institute grant, FQXi-RFP-2019-IAF, and by an NSERC Discovery Grant.


\begin{thebibliography}{44}%
\makeatletter
\providecommand \@ifxundefined [1]{%
 \@ifx{#1\undefined}
}%
\providecommand \@ifnum [1]{%
 \ifnum #1\expandafter \@firstoftwo
 \else \expandafter \@secondoftwo
 \fi
}%
\providecommand \@ifx [1]{%
 \ifx #1\expandafter \@firstoftwo
 \else \expandafter \@secondoftwo
 \fi
}%
\providecommand \natexlab [1]{#1}%
\providecommand \enquote  [1]{``#1''}%
\providecommand \bibnamefont  [1]{#1}%
\providecommand \bibfnamefont [1]{#1}%
\providecommand \citenamefont [1]{#1}%
\providecommand \href@noop [0]{\@secondoftwo}%
\providecommand \href [0]{\begingroup \@sanitize@url \@href}%
\providecommand \@href[1]{\@@startlink{#1}\@@href}%
\providecommand \@@href[1]{\endgroup#1\@@endlink}%
\providecommand \@sanitize@url [0]{\catcode `\\12\catcode `\$12\catcode
  `\&12\catcode `\#12\catcode `\^12\catcode `\_12\catcode `\%12\relax}%
\providecommand \@@startlink[1]{}%
\providecommand \@@endlink[0]{}%
\providecommand \url  [0]{\begingroup\@sanitize@url \@url }%
\providecommand \@url [1]{\endgroup\@href {#1}{\urlprefix }}%
\providecommand \urlprefix  [0]{URL }%
\providecommand \Eprint [0]{\href }%
\providecommand \doibase [0]{https://doi.org/}%
\providecommand \selectlanguage [0]{\@gobble}%
\providecommand \bibinfo  [0]{\@secondoftwo}%
\providecommand \bibfield  [0]{\@secondoftwo}%
\providecommand \translation [1]{[#1]}%
\providecommand \BibitemOpen [0]{}%
\providecommand \bibitemStop [0]{}%
\providecommand \bibitemNoStop [0]{.\EOS\space}%
\providecommand \EOS [0]{\spacefactor3000\relax}%
\providecommand \BibitemShut  [1]{\csname bibitem#1\endcsname}%
\let\auto@bib@innerbib\@empty
\bibitem [{\citenamefont {Landauer}(1961)}]{landauer1961irreversibility}%
  \BibitemOpen
  \bibfield  {author} {\bibinfo {author} {\bibfnamefont {R.}~\bibnamefont
  {Landauer}},\ }\bibfield  {title} {\bibinfo {title} {Irreversibility and heat
  generation in the computing process},\ }\href
  {https://doi.org/10.1147/rd.53.0183} {\bibfield  {journal} {\bibinfo
  {journal} {IBM J. Res. Develop.}\ }\textbf {\bibinfo {volume} {5}},\ \bibinfo
  {pages} {183} (\bibinfo {year} {1961})}\BibitemShut {NoStop}%
\bibitem [{\citenamefont {Parrondo}\ \emph {et~al.}(2015)\citenamefont
  {Parrondo}, \citenamefont {Horowitz},\ and\ \citenamefont
  {Sagawa}}]{Parrondo2015_infoTD}%
  \BibitemOpen
  \bibfield  {author} {\bibinfo {author} {\bibfnamefont {J.~M.~R.}\
  \bibnamefont {Parrondo}}, \bibinfo {author} {\bibfnamefont {J.~M.}\
  \bibnamefont {Horowitz}},\ and\ \bibinfo {author} {\bibfnamefont
  {T.}~\bibnamefont {Sagawa}},\ }\bibfield  {title} {\bibinfo {title}
  {Thermodynamics of information},\ }\href {https://doi.org/10.1038/NPHYS3230}
  {\bibfield  {journal} {\bibinfo  {journal} {Nature Phys.}\ }\textbf {\bibinfo
  {volume} {11}},\ \bibinfo {pages} {131} (\bibinfo {year} {2015})}\BibitemShut
  {NoStop}%
\bibitem [{\citenamefont {Leff}\ and\ \citenamefont
  {Rex}(1990)}]{Leff19903MaxwellsDemon}%
  \BibitemOpen
  \bibinfo {editor} {\bibfnamefont {H.~S.}\ \bibnamefont {Leff}}\ and\ \bibinfo
  {editor} {\bibfnamefont {A.~F.}\ \bibnamefont {Rex}},\ eds.,\ \href@noop {}
  {\emph {\bibinfo {title} {Maxwell's Demon: Entropy, Information,
  Computing}}}\ (\bibinfo  {publisher} {Princeton University Press},\ \bibinfo
  {address} {Princeton, NJ},\ \bibinfo {year} {1990})\BibitemShut {NoStop}%
\bibitem [{\citenamefont {Rex}(2017)}]{Rex2017MaxwellsDemon}%
  \BibitemOpen
  \bibfield  {author} {\bibinfo {author} {\bibfnamefont {A.}~\bibnamefont
  {Rex}},\ }\bibfield  {title} {\bibinfo {title} {{Maxwell's demon---A
  historical review}},\ }\href {https://doi.org/10.3390/e19060240} {\bibfield
  {journal} {\bibinfo  {journal} {Entropy}\ }\textbf {\bibinfo {volume} {19}},\
  \bibinfo {pages} {240} (\bibinfo {year} {2017})}\BibitemShut {NoStop}%
\bibitem [{\citenamefont {B{\'e}rut}\ \emph {et~al.}(2012)\citenamefont
  {B{\'e}rut}, \citenamefont {Arakelyan}, \citenamefont {Petrosyan},
  \citenamefont {Ciliberto}, \citenamefont {Dillenschneider},\ and\
  \citenamefont {Lutz}}]{berut2012experimental}%
  \BibitemOpen
  \bibfield  {author} {\bibinfo {author} {\bibfnamefont {A.}~\bibnamefont
  {B{\'e}rut}}, \bibinfo {author} {\bibfnamefont {A.}~\bibnamefont
  {Arakelyan}}, \bibinfo {author} {\bibfnamefont {A.}~\bibnamefont
  {Petrosyan}}, \bibinfo {author} {\bibfnamefont {S.}~\bibnamefont
  {Ciliberto}}, \bibinfo {author} {\bibfnamefont {R.}~\bibnamefont
  {Dillenschneider}},\ and\ \bibinfo {author} {\bibfnamefont {E.}~\bibnamefont
  {Lutz}},\ }\bibfield  {title} {\bibinfo {title} {Experimental verification of
  {L}andauer's principle linking information and thermodynamics},\ }\href
  {https://doi.org/10.1038/nature10872} {\bibfield  {journal} {\bibinfo
  {journal} {Nature}\ }\textbf {\bibinfo {volume} {483}},\ \bibinfo {pages}
  {187} (\bibinfo {year} {2012})}\BibitemShut {NoStop}%
\bibitem [{\citenamefont {Jun}\ \emph {et~al.}(2014)\citenamefont {Jun},
  \citenamefont {Gavrilov},\ and\ \citenamefont {Bechhoefer}}]{jun2014high}%
  \BibitemOpen
  \bibfield  {author} {\bibinfo {author} {\bibfnamefont {Y.}~\bibnamefont
  {Jun}}, \bibinfo {author} {\bibfnamefont {M.}~\bibnamefont {Gavrilov}},\ and\
  \bibinfo {author} {\bibfnamefont {J.}~\bibnamefont {Bechhoefer}},\ }\bibfield
   {title} {\bibinfo {title} {{High-Precision Test of Landauer’s Principle in
  a Feedback Trap}},\ }\href {https://doi.org/10.1103/PhysRevLett.113.190601}
  {\bibfield  {journal} {\bibinfo  {journal} {Phys. Rev. Lett.}\ }\textbf
  {\bibinfo {volume} {113}},\ \bibinfo {pages} {190601} (\bibinfo {year}
  {2014})}\BibitemShut {NoStop}%
\bibitem [{\citenamefont {Gavrilov}\ and\ \citenamefont
  {Bechhoefer}(2016)}]{gavrilov2016erasure}%
  \BibitemOpen
  \bibfield  {author} {\bibinfo {author} {\bibfnamefont {M.}~\bibnamefont
  {Gavrilov}}\ and\ \bibinfo {author} {\bibfnamefont {J.}~\bibnamefont
  {Bechhoefer}},\ }\bibfield  {title} {\bibinfo {title} {{Erasure without Work
  in an Asymmetric Double-Well Potential}},\ }\href
  {https://doi.org/10.1103/PhysRevLett.117.200601} {\bibfield  {journal}
  {\bibinfo  {journal} {Phys. Rev. Lett.}\ }\textbf {\bibinfo {volume} {117}},\
  \bibinfo {pages} {200601} (\bibinfo {year} {2016})}\BibitemShut {NoStop}%
\bibitem [{\citenamefont {Gavrilov}\ \emph {et~al.}(2017)\citenamefont
  {Gavrilov}, \citenamefont {Ch{\'e}trite},\ and\ \citenamefont
  {Bechhoefer}}]{gavrilov2017direct}%
  \BibitemOpen
  \bibfield  {author} {\bibinfo {author} {\bibfnamefont {M.}~\bibnamefont
  {Gavrilov}}, \bibinfo {author} {\bibfnamefont {R.}~\bibnamefont
  {Ch{\'e}trite}},\ and\ \bibinfo {author} {\bibfnamefont {J.}~\bibnamefont
  {Bechhoefer}},\ }\bibfield  {title} {\bibinfo {title} {Direct measurement of
  weakly nonequilibrium system entropy is consistent with {G}ibbs-{S}hannon
  form},\ }\href {https://doi.org/10.1073/pnas.1708689114} {\bibfield
  {journal} {\bibinfo  {journal} {Proc. Natl. Acad. Sci. U.S.A.}\ }\textbf
  {\bibinfo {volume} {114}},\ \bibinfo {pages} {11097} (\bibinfo {year}
  {2017})}\BibitemShut {NoStop}%
\bibitem [{\citenamefont {Saira}\ \emph {et~al.}(2020)\citenamefont {Saira},
  \citenamefont {Matheny}, \citenamefont {Katti}, \citenamefont {Fon},
  \citenamefont {Wimsatt}, \citenamefont {Crutchfield}, \citenamefont {Han},\
  and\ \citenamefont {Roukes}}]{saira2020nonequilibrium}%
  \BibitemOpen
  \bibfield  {author} {\bibinfo {author} {\bibfnamefont {O.-P.}\ \bibnamefont
  {Saira}}, \bibinfo {author} {\bibfnamefont {M.~H.}\ \bibnamefont {Matheny}},
  \bibinfo {author} {\bibfnamefont {R.}~\bibnamefont {Katti}}, \bibinfo
  {author} {\bibfnamefont {W.}~\bibnamefont {Fon}}, \bibinfo {author}
  {\bibfnamefont {G.}~\bibnamefont {Wimsatt}}, \bibinfo {author} {\bibfnamefont
  {J.~P.}\ \bibnamefont {Crutchfield}}, \bibinfo {author} {\bibfnamefont
  {S.}~\bibnamefont {Han}},\ and\ \bibinfo {author} {\bibfnamefont {M.~L.}\
  \bibnamefont {Roukes}},\ }\bibfield  {title} {\bibinfo {title}
  {Nonequilibrium thermodynamics of erasure with superconducting flux logic},\
  }\href {https://doi.org/10.1103/PhysRevResearch.2.013249} {\bibfield
  {journal} {\bibinfo  {journal} {Phys. Rev. Research}\ }\textbf {\bibinfo
  {volume} {2}},\ \bibinfo {pages} {013249} (\bibinfo {year}
  {2020})}\BibitemShut {NoStop}%
\bibitem [{\citenamefont {Zulkowski}\ and\ \citenamefont
  {DeWeese}(2014)}]{zulkowski2014optimal}%
  \BibitemOpen
  \bibfield  {author} {\bibinfo {author} {\bibfnamefont {P.~R.}\ \bibnamefont
  {Zulkowski}}\ and\ \bibinfo {author} {\bibfnamefont {M.~R.}\ \bibnamefont
  {DeWeese}},\ }\bibfield  {title} {\bibinfo {title} {Optimal finite-time
  erasure of a classical bit},\ }\href
  {https://doi.org/10.1103/PhysRevE.89.052140} {\bibfield  {journal} {\bibinfo
  {journal} {Phys. Rev. E}\ }\textbf {\bibinfo {volume} {89}},\ \bibinfo
  {pages} {052140} (\bibinfo {year} {2014})}\BibitemShut {NoStop}%
\bibitem [{\citenamefont {Zulkowski}\ and\ \citenamefont
  {DeWeese}(2015)}]{zulkowski2015optimal}%
  \BibitemOpen
  \bibfield  {author} {\bibinfo {author} {\bibfnamefont {P.~R.}\ \bibnamefont
  {Zulkowski}}\ and\ \bibinfo {author} {\bibfnamefont {M.~R.}\ \bibnamefont
  {DeWeese}},\ }\bibfield  {title} {\bibinfo {title} {Optimal control of
  overdamped systems},\ }\href {https://doi.org/10.1103/PhysRevE.92.032117 I.}
  {\bibfield  {journal} {\bibinfo  {journal} {Phys. Rev. E}\ }\textbf {\bibinfo
  {volume} {92}},\ \bibinfo {pages} {032117} (\bibinfo {year}
  {2015})}\BibitemShut {NoStop}%
\bibitem [{\citenamefont {Boyd}\ \emph
  {et~al.}(2018{\natexlab{a}})\citenamefont {Boyd}, \citenamefont {Patra},
  \citenamefont {Jarzynski},\ and\ \citenamefont
  {Crutchfield}}]{boyd2018shortcuts}%
  \BibitemOpen
  \bibfield  {author} {\bibinfo {author} {\bibfnamefont {A.~B.}\ \bibnamefont
  {Boyd}}, \bibinfo {author} {\bibfnamefont {A.}~\bibnamefont {Patra}},
  \bibinfo {author} {\bibfnamefont {C.}~\bibnamefont {Jarzynski}},\ and\
  \bibinfo {author} {\bibfnamefont {J.~P.}\ \bibnamefont {Crutchfield}},\
  }\bibfield  {title} {\bibinfo {title} {Shortcuts to thermodynamic computing:
  The cost of fast and faithful erasure},\ }\href
  {https://arxiv.org/abs/1812.11241} {\bibfield  {journal} {\bibinfo  {journal}
  {arXiv:1812.11241}\ } (\bibinfo {year} {2018}{\natexlab{a}})}\BibitemShut
  {NoStop}%
\bibitem [{\citenamefont {Andresen}(2011)}]{andresen2011current}%
  \BibitemOpen
  \bibfield  {author} {\bibinfo {author} {\bibfnamefont {B.}~\bibnamefont
  {Andresen}},\ }\bibfield  {title} {\bibinfo {title} {Current trends in
  finite-time thermodynamics},\ }\href {https://doi.org/0.1002/anie.201001411}
  {\bibfield  {journal} {\bibinfo  {journal} {Angew. Chem. Int. Ed.}\ }\textbf
  {\bibinfo {volume} {50}},\ \bibinfo {pages} {2690} (\bibinfo {year}
  {2011})}\BibitemShut {NoStop}%
\bibitem [{\citenamefont {Seifert}(2012)}]{seifert2012stochastic}%
  \BibitemOpen
  \bibfield  {author} {\bibinfo {author} {\bibfnamefont {U.}~\bibnamefont
  {Seifert}},\ }\bibfield  {title} {\bibinfo {title} {Stochastic
  thermodynamics, fluctuation theorems and molecular machines},\ }\href
  {https://doi.org/10.1088/0034-4885/75/12/126001} {\bibfield  {journal}
  {\bibinfo  {journal} {Rep. Prog. Phys.}\ }\textbf {\bibinfo {volume} {75}},\
  \bibinfo {pages} {126001} (\bibinfo {year} {2012})}\BibitemShut {NoStop}%
\bibitem [{\citenamefont {Van~den Broeck}\ and\ \citenamefont
  {Esposito}(2015)}]{van2015ensemble}%
  \BibitemOpen
  \bibfield  {author} {\bibinfo {author} {\bibfnamefont {C.}~\bibnamefont
  {Van~den Broeck}}\ and\ \bibinfo {author} {\bibfnamefont {M.}~\bibnamefont
  {Esposito}},\ }\bibfield  {title} {\bibinfo {title} {Ensemble and trajectory
  thermodynamics: {A} brief introduction},\ }\href
  {https://doi.org/10.1016/j.physa.2014.04.035} {\bibfield  {journal} {\bibinfo
   {journal} {Physica A}\ }\textbf {\bibinfo {volume} {418}},\ \bibinfo {pages}
  {6} (\bibinfo {year} {2015})}\BibitemShut {NoStop}%
\bibitem [{\citenamefont {Schmiedl}\ and\ \citenamefont
  {Seifert}(2007)}]{schmiedl07}%
  \BibitemOpen
  \bibfield  {author} {\bibinfo {author} {\bibfnamefont {T.}~\bibnamefont
  {Schmiedl}}\ and\ \bibinfo {author} {\bibfnamefont {U.}~\bibnamefont
  {Seifert}},\ }\bibfield  {title} {\bibinfo {title} {{Optimal Finite-Time
  Processes In Stochastic Thermodynamics}},\ }\href
  {https://doi.org/10.1103/PhysRevLett.98.108301} {\bibfield  {journal}
  {\bibinfo  {journal} {Phys. Rev. Lett.}\ }\textbf {\bibinfo {volume} {98}},\
  \bibinfo {pages} {108301} (\bibinfo {year} {2007})}\BibitemShut {NoStop}%
\bibitem [{\citenamefont {Bonan{\c{c}}a}\ and\ \citenamefont
  {Deffner}(2014)}]{bonancca2014optimal}%
  \BibitemOpen
  \bibfield  {author} {\bibinfo {author} {\bibfnamefont {M.~V.}\ \bibnamefont
  {Bonan{\c{c}}a}}\ and\ \bibinfo {author} {\bibfnamefont {S.}~\bibnamefont
  {Deffner}},\ }\bibfield  {title} {\bibinfo {title} {Optimal driving of
  isothermal processes close to equilibrium},\ }\href
  {https://doi.org/10.1063/1.4885277} {\bibfield  {journal} {\bibinfo
  {journal} {J. Chem. Phys.}\ }\textbf {\bibinfo {volume} {140}},\ \bibinfo
  {pages} {244119} (\bibinfo {year} {2014})}\BibitemShut {NoStop}%
\bibitem [{\citenamefont {Sivak}\ and\ \citenamefont
  {Crooks}(2012)}]{Sivak2012TDMetrics}%
  \BibitemOpen
  \bibfield  {author} {\bibinfo {author} {\bibfnamefont {D.~A.}\ \bibnamefont
  {Sivak}}\ and\ \bibinfo {author} {\bibfnamefont {G.~E.}\ \bibnamefont
  {Crooks}},\ }\bibfield  {title} {\bibinfo {title} {{Thermodynamic metrics and
  optimal paths}},\ }\href {https://doi.org/10.1103/PhysRevLett.108.190602}
  {\bibfield  {journal} {\bibinfo  {journal} {Phys. Rev. Lett.}\ }\textbf
  {\bibinfo {volume} {108}},\ \bibinfo {pages} {190602} (\bibinfo {year}
  {2012})}\BibitemShut {NoStop}%
\bibitem [{\citenamefont {Tafoya}\ \emph {et~al.}(2019)\citenamefont {Tafoya},
  \citenamefont {Large}, \citenamefont {Liu}, \citenamefont {Bustamante},\ and\
  \citenamefont {Sivak}}]{tafoya2019using}%
  \BibitemOpen
  \bibfield  {author} {\bibinfo {author} {\bibfnamefont {S.}~\bibnamefont
  {Tafoya}}, \bibinfo {author} {\bibfnamefont {S.~J.}\ \bibnamefont {Large}},
  \bibinfo {author} {\bibfnamefont {S.}~\bibnamefont {Liu}}, \bibinfo {author}
  {\bibfnamefont {C.}~\bibnamefont {Bustamante}},\ and\ \bibinfo {author}
  {\bibfnamefont {D.~A.}\ \bibnamefont {Sivak}},\ }\bibfield  {title} {\bibinfo
  {title} {Using a system's equilibrium behavior to reduce its energy
  dissipation in nonequilibrium processes},\ }\href
  {https://doi.org/10.1073/pnas.1817778116} {\bibfield  {journal} {\bibinfo
  {journal} {Proc. Natl. Acad. Sci. U.S.A.}\ }\textbf {\bibinfo {volume}
  {116}},\ \bibinfo {pages} {5920} (\bibinfo {year} {2019})}\BibitemShut
  {NoStop}%
\bibitem [{\citenamefont {Plata}\ \emph {et~al.}(2020)\citenamefont {Plata},
  \citenamefont {Gu{\'e}ry-Odelin}, \citenamefont {Trizac},\ and\ \citenamefont
  {Prados}}]{plata2020finite}%
  \BibitemOpen
  \bibfield  {author} {\bibinfo {author} {\bibfnamefont {C.~A.}\ \bibnamefont
  {Plata}}, \bibinfo {author} {\bibfnamefont {D.}~\bibnamefont
  {Gu{\'e}ry-Odelin}}, \bibinfo {author} {\bibfnamefont {E.}~\bibnamefont
  {Trizac}},\ and\ \bibinfo {author} {\bibfnamefont {A.}~\bibnamefont
  {Prados}},\ }\bibfield  {title} {\bibinfo {title} {Finite-time adiabatic
  processes: Derivation and speed limit},\ }\href
  {https://doi.org/10.1103/PhysRevE.101.032129} {\bibfield  {journal} {\bibinfo
   {journal} {Phys. Rev. E}\ }\textbf {\bibinfo {volume} {101}},\ \bibinfo
  {pages} {032129} (\bibinfo {year} {2020})}\BibitemShut {NoStop}%
\bibitem [{\citenamefont {Bryant}\ and\ \citenamefont
  {Machta}(2020)}]{bryant2020energy}%
  \BibitemOpen
  \bibfield  {author} {\bibinfo {author} {\bibfnamefont {S.~J.}\ \bibnamefont
  {Bryant}}\ and\ \bibinfo {author} {\bibfnamefont {B.~B.}\ \bibnamefont
  {Machta}},\ }\bibfield  {title} {\bibinfo {title} {Energy dissipation bounds
  for autonomous thermodynamic cycles},\ }\href
  {https://doi.org/10.1073/pnas.1915676117} {\bibfield  {journal} {\bibinfo
  {journal} {Proc. Natl. Acad. Sci. U.S.A.}\ }\textbf {\bibinfo {volume}
  {117}},\ \bibinfo {pages} {3478} (\bibinfo {year} {2020})}\BibitemShut
  {NoStop}%
\bibitem [{\citenamefont {Diana}\ \emph {et~al.}(2013)\citenamefont {Diana},
  \citenamefont {Bagci},\ and\ \citenamefont {Esposito}}]{diana2013finite}%
  \BibitemOpen
  \bibfield  {author} {\bibinfo {author} {\bibfnamefont {G.}~\bibnamefont
  {Diana}}, \bibinfo {author} {\bibfnamefont {G.~B.}\ \bibnamefont {Bagci}},\
  and\ \bibinfo {author} {\bibfnamefont {M.}~\bibnamefont {Esposito}},\
  }\bibfield  {title} {\bibinfo {title} {Finite-time erasing of information
  stored in fermionic bits},\ }\href
  {https://doi.org/10.1103/PhysRevE.87.012111} {\bibfield  {journal} {\bibinfo
  {journal} {Phys. Rev. E}\ }\textbf {\bibinfo {volume} {87}},\ \bibinfo
  {pages} {012111} (\bibinfo {year} {2013})}\BibitemShut {NoStop}%
\bibitem [{\citenamefont {Boyd}\ \emph
  {et~al.}(2018{\natexlab{b}})\citenamefont {Boyd}, \citenamefont {Mandal},\
  and\ \citenamefont {Crutchfield}}]{boyd2018thermodynamics}%
  \BibitemOpen
  \bibfield  {author} {\bibinfo {author} {\bibfnamefont {A.~B.}\ \bibnamefont
  {Boyd}}, \bibinfo {author} {\bibfnamefont {D.}~\bibnamefont {Mandal}},\ and\
  \bibinfo {author} {\bibfnamefont {J.~P.}\ \bibnamefont {Crutchfield}},\
  }\bibfield  {title} {\bibinfo {title} {Thermodynamics of modularity:
  {S}tructural costs beyond the {L}andauer bound},\ }\href
  {https://doi.org/10.1103/PhysRevX.8.031036} {\bibfield  {journal} {\bibinfo
  {journal} {Phys. Rev. X}\ }\textbf {\bibinfo {volume} {8}},\ \bibinfo {pages}
  {031036} (\bibinfo {year} {2018}{\natexlab{b}})}\BibitemShut {NoStop}%
\bibitem [{\citenamefont {Riechers}\ \emph {et~al.}(2019)\citenamefont
  {Riechers}, \citenamefont {Boyd}, \citenamefont {Wimsatt},\ and\
  \citenamefont {Crutchfield}}]{riechers2019balancing}%
  \BibitemOpen
  \bibfield  {author} {\bibinfo {author} {\bibfnamefont {P.~M.}\ \bibnamefont
  {Riechers}}, \bibinfo {author} {\bibfnamefont {A.~B.}\ \bibnamefont {Boyd}},
  \bibinfo {author} {\bibfnamefont {G.~W.}\ \bibnamefont {Wimsatt}},\ and\
  \bibinfo {author} {\bibfnamefont {J.~P.}\ \bibnamefont {Crutchfield}},\
  }\bibfield  {title} {\bibinfo {title} {{Balancing Error and Dissipation in
  Highly-Reliable Computing}},\ }\href {https://arxiv.org/abs/1909.06650}
  {\bibfield  {journal} {\bibinfo  {journal} {arXiv:1909.06650}\ } (\bibinfo
  {year} {2019})}\BibitemShut {NoStop}%
\bibitem [{\citenamefont {Aurell}\ \emph {et~al.}(2011)\citenamefont {Aurell},
  \citenamefont {Mej{\'\i}a-Monasterio},\ and\ \citenamefont
  {Muratore-Ginanneschi}}]{aurell2011optimal}%
  \BibitemOpen
  \bibfield  {author} {\bibinfo {author} {\bibfnamefont {E.}~\bibnamefont
  {Aurell}}, \bibinfo {author} {\bibfnamefont {C.}~\bibnamefont
  {Mej{\'\i}a-Monasterio}},\ and\ \bibinfo {author} {\bibfnamefont
  {P.}~\bibnamefont {Muratore-Ginanneschi}},\ }\bibfield  {title} {\bibinfo
  {title} {{Optimal Protocols and Optimal Transport in Stochastic
  Thermodynamics}},\ }\href {https://doi.org/10.1103/PhysRevLett.106.250601}
  {\bibfield  {journal} {\bibinfo  {journal} {Phys. Rev. Lett.}\ }\textbf
  {\bibinfo {volume} {106}},\ \bibinfo {pages} {250601} (\bibinfo {year}
  {2011})}\BibitemShut {NoStop}%
\bibitem [{\citenamefont {Aurell}\ \emph {et~al.}(2012)\citenamefont {Aurell},
  \citenamefont {Gawedzki}, \citenamefont {Mejia-Monasterio}, \citenamefont
  {Mohayaee},\ and\ \citenamefont {Muratore-Ginanneschi}}]{aurell2012refined}%
  \BibitemOpen
  \bibfield  {author} {\bibinfo {author} {\bibfnamefont {E.}~\bibnamefont
  {Aurell}}, \bibinfo {author} {\bibfnamefont {K.}~\bibnamefont {Gawedzki}},
  \bibinfo {author} {\bibfnamefont {C.}~\bibnamefont {Mejia-Monasterio}},
  \bibinfo {author} {\bibfnamefont {R.}~\bibnamefont {Mohayaee}},\ and\
  \bibinfo {author} {\bibfnamefont {P.}~\bibnamefont {Muratore-Ginanneschi}},\
  }\bibfield  {title} {\bibinfo {title} {Refined second law of thermodynamics
  for fast random processes},\ }\href
  {https://doi.org/10.1007/s10955-012-0478-x} {\bibfield  {journal} {\bibinfo
  {journal} {J. Stat. Phys.}\ }\textbf {\bibinfo {volume} {147}},\ \bibinfo
  {pages} {487} (\bibinfo {year} {2012})}\BibitemShut {NoStop}%
\bibitem [{\citenamefont {Proesmans}\ \emph {et~al.}(2020)\citenamefont
  {Proesmans}, \citenamefont {Ehrich},\ and\ \citenamefont
  {Bechhoefer}}]{proesmans20}%
  \BibitemOpen
  \bibfield  {author} {\bibinfo {author} {\bibfnamefont {K.}~\bibnamefont
  {Proesmans}}, \bibinfo {author} {\bibfnamefont {J.}~\bibnamefont {Ehrich}},\
  and\ \bibinfo {author} {\bibfnamefont {J.}~\bibnamefont {Bechhoefer}},\
  }\bibfield  {title} {\bibinfo {title} {Optimal finite-time bit erasure under
  full control},\ }\href@noop {} {\bibfield  {journal} {\bibinfo  {journal}
  {Accompanying manuscript}\ } (\bibinfo {year} {2020})}\BibitemShut {NoStop}%
\bibitem [{\citenamefont {Villani}(2003)}]{villani2003topics}%
  \BibitemOpen
  \bibfield  {author} {\bibinfo {author} {\bibfnamefont {C.}~\bibnamefont
  {Villani}},\ }\href {https://doi.org/10.1090/gsm/058} {\emph {\bibinfo
  {title} {Topics in optimal transportation}}},\ \bibinfo {number} {58}\
  (\bibinfo  {publisher} {American Mathematical Soc.},\ \bibinfo {year}
  {2003})\BibitemShut {NoStop}%
\bibitem [{\citenamefont {L{\'e}vy}\ and\ \citenamefont
  {Schwindt}(2018)}]{levy2018notions}%
  \BibitemOpen
  \bibfield  {author} {\bibinfo {author} {\bibfnamefont {B.}~\bibnamefont
  {L{\'e}vy}}\ and\ \bibinfo {author} {\bibfnamefont {E.~L.}\ \bibnamefont
  {Schwindt}},\ }\bibfield  {title} {\bibinfo {title} {Notions of optimal
  transport theory and how to implement them on a computer},\ }\href
  {https://doi.org/10.1016/j.cag.2018.01.009} {\bibfield  {journal} {\bibinfo
  {journal} {Computers \& Graphics}\ }\textbf {\bibinfo {volume} {72}},\
  \bibinfo {pages} {135} (\bibinfo {year} {2018})}\BibitemShut {NoStop}%
\bibitem [{\citenamefont {Zhang}(2019)}]{zhang2019work}%
  \BibitemOpen
  \bibfield  {author} {\bibinfo {author} {\bibfnamefont {Y.}~\bibnamefont
  {Zhang}},\ }\bibfield  {title} {\bibinfo {title} {Work needed to drive a
  thermodynamic system between two distributions},\ }\href
  {https://doi.org/10.1209/0295-5075/128/30002} {\bibfield  {journal} {\bibinfo
   {journal} {Europhys. Lett.}\ }\textbf {\bibinfo {volume} {128}},\ \bibinfo
  {pages} {30002} (\bibinfo {year} {2019})}\BibitemShut {NoStop}%
\bibitem [{\citenamefont {Zhang}(2020)}]{zhang2020optimization}%
  \BibitemOpen
  \bibfield  {author} {\bibinfo {author} {\bibfnamefont {Y.}~\bibnamefont
  {Zhang}},\ }\bibfield  {title} {\bibinfo {title} {{Optimization of Stochastic
  Thermodynamic Machines}},\ }\href
  {https://doi.org/10.1007/s10955-020-02508-0} {\bibfield  {journal} {\bibinfo
  {journal} {J. Stat. Phys.}\ }\textbf {\bibinfo {volume} {178}},\ \bibinfo
  {pages} {1336} (\bibinfo {year} {2020})}\BibitemShut {NoStop}%
\bibitem [{\citenamefont {Esposito}\ and\ \citenamefont {Van~den
  Broeck}(2011)}]{Esposito2011NonequilibriumFreeEnergy}%
  \BibitemOpen
  \bibfield  {author} {\bibinfo {author} {\bibfnamefont {M.}~\bibnamefont
  {Esposito}}\ and\ \bibinfo {author} {\bibfnamefont {C.}~\bibnamefont {Van~den
  Broeck}},\ }\bibfield  {title} {\bibinfo {title} {Second law and {L}andauer
  principle far from equilibrium},\ }\href
  {https://doi.org/10.1209/0295-5075/95/40004 August 2011 www.epljournal.org
  Second} {\bibfield  {journal} {\bibinfo  {journal} {Europhys. Lett.}\
  }\textbf {\bibinfo {volume} {95}},\ \bibinfo {pages} {40004} (\bibinfo {year}
  {2011})}\BibitemShut {NoStop}%
\bibitem [{\citenamefont {Shiraishi}\ \emph {et~al.}(2018)\citenamefont
  {Shiraishi}, \citenamefont {Funo},\ and\ \citenamefont
  {Saito}}]{shiraishi2018speed}%
  \BibitemOpen
  \bibfield  {author} {\bibinfo {author} {\bibfnamefont {N.}~\bibnamefont
  {Shiraishi}}, \bibinfo {author} {\bibfnamefont {K.}~\bibnamefont {Funo}},\
  and\ \bibinfo {author} {\bibfnamefont {K.}~\bibnamefont {Saito}},\ }\bibfield
   {title} {\bibinfo {title} {Speed limit for classical stochastic processes},\
  }\href {https://doi.org/10.1103/PhysRevLett.121.070601} {\bibfield  {journal}
  {\bibinfo  {journal} {Phys. Rev. Lett.}\ }\textbf {\bibinfo {volume} {121}},\
  \bibinfo {pages} {070601} (\bibinfo {year} {2018})}\BibitemShut {NoStop}%
\bibitem [{\citenamefont {Zhang}(2018)}]{zhang2018comment}%
  \BibitemOpen
  \bibfield  {author} {\bibinfo {author} {\bibfnamefont {Y.}~\bibnamefont
  {Zhang}},\ }\bibfield  {title} {\bibinfo {title} {Comment on ``{S}peed limit
  for classical stochastic processes"},\ }\href
  {https://arxiv.org/abs/1811.06978} {\bibfield  {journal} {\bibinfo  {journal}
  {arXiv:1811.06978}\ } (\bibinfo {year} {2018})}\BibitemShut {NoStop}%
\bibitem [{\citenamefont {Dechant}\ and\ \citenamefont
  {Sakurai}(2019)}]{dechant2019thermodynamic}%
  \BibitemOpen
  \bibfield  {author} {\bibinfo {author} {\bibfnamefont {A.}~\bibnamefont
  {Dechant}}\ and\ \bibinfo {author} {\bibfnamefont {Y.}~\bibnamefont
  {Sakurai}},\ }\bibfield  {title} {\bibinfo {title} {Thermodynamic
  interpretation of {W}asserstein distance},\ }\href
  {https://arxiv.org/abs/1912.08405} {\bibfield  {journal} {\bibinfo  {journal}
  {arXiv:1912.08405}\ } (\bibinfo {year} {2019})}\BibitemShut {NoStop}%
\bibitem [{\citenamefont {B{\'e}rut}\ \emph {et~al.}(2013)\citenamefont
  {B{\'e}rut}, \citenamefont {Petrosyan},\ and\ \citenamefont
  {Ciliberto}}]{berut2013detailed}%
  \BibitemOpen
  \bibfield  {author} {\bibinfo {author} {\bibfnamefont {A.}~\bibnamefont
  {B{\'e}rut}}, \bibinfo {author} {\bibfnamefont {A.}~\bibnamefont
  {Petrosyan}},\ and\ \bibinfo {author} {\bibfnamefont {S.}~\bibnamefont
  {Ciliberto}},\ }\bibfield  {title} {\bibinfo {title} {Detailed {J}arzynski
  equality applied to a logically irreversible procedure},\ }\href
  {https://doi.org/10.1209/0295-5075/103/60002 September} {\bibfield  {journal}
  {\bibinfo  {journal} {Europhys. Lett.}\ }\textbf {\bibinfo {volume} {103}},\
  \bibinfo {pages} {60002} (\bibinfo {year} {2013})}\BibitemShut {NoStop}%
\bibitem [{\citenamefont {B{\'e}rut}\ \emph {et~al.}(2015)\citenamefont
  {B{\'e}rut}, \citenamefont {Petrosyan},\ and\ \citenamefont
  {Ciliberto}}]{berut2015information}%
  \BibitemOpen
  \bibfield  {author} {\bibinfo {author} {\bibfnamefont {A.}~\bibnamefont
  {B{\'e}rut}}, \bibinfo {author} {\bibfnamefont {A.}~\bibnamefont
  {Petrosyan}},\ and\ \bibinfo {author} {\bibfnamefont {S.}~\bibnamefont
  {Ciliberto}},\ }\bibfield  {title} {\bibinfo {title} {Information and
  thermodynamics: experimental verification of {L}andauer's {E}rasure
  principle},\ }\href {https://doi.org/10.1088/1742-5468/2015/06/P06015}
  {\bibfield  {journal} {\bibinfo  {journal} {J. Stat. Mech.}\ }\textbf
  {\bibinfo {volume} {2015}},\ \bibinfo {pages} {P06015} (\bibinfo {year}
  {2015})}\BibitemShut {NoStop}%
\bibitem [{Pro()}]{Proesmans20SI}%
  \BibitemOpen
  \href@noop {} {}\bibinfo {note} {See Supplemental Material at [link] for
  discussion of experimental setups and parameter values, along with
  accompanying calculations on the various bounds.}\BibitemShut {Stop}%
\bibitem [{\citenamefont {Deshpande}\ \emph {et~al.}(2017)\citenamefont
  {Deshpande}, \citenamefont {Gopalkrishnan}, \citenamefont {Ouldridge},\ and\
  \citenamefont {Jones}}]{deshpande2017designing}%
  \BibitemOpen
  \bibfield  {author} {\bibinfo {author} {\bibfnamefont {A.}~\bibnamefont
  {Deshpande}}, \bibinfo {author} {\bibfnamefont {M.}~\bibnamefont
  {Gopalkrishnan}}, \bibinfo {author} {\bibfnamefont {T.~E.}\ \bibnamefont
  {Ouldridge}},\ and\ \bibinfo {author} {\bibfnamefont {N.~S.}\ \bibnamefont
  {Jones}},\ }\bibfield  {title} {\bibinfo {title} {Designing the optimal bit:
  balancing energetic cost, speed and reliability},\ }\href
  {https://doi.org/10.1098/rspa.2017.0117} {\bibfield  {journal} {\bibinfo
  {journal} {Proc. R. Soc. Lond.}\ }\textbf {\bibinfo {volume} {473}},\
  \bibinfo {pages} {20170117} (\bibinfo {year} {2017})}\BibitemShut {NoStop}%
\bibitem [{\citenamefont {Reeb}\ and\ \citenamefont
  {Wolf}(2014)}]{reeb2014improved}%
  \BibitemOpen
  \bibfield  {author} {\bibinfo {author} {\bibfnamefont {D.}~\bibnamefont
  {Reeb}}\ and\ \bibinfo {author} {\bibfnamefont {M.~M.}\ \bibnamefont
  {Wolf}},\ }\bibfield  {title} {\bibinfo {title} {An improved {L}andauer
  principle with finite-size corrections},\ }\href
  {https://doi.org/10.1088/1367-2630/16/10/103011} {\bibfield  {journal}
  {\bibinfo  {journal} {New J. Phys.}\ }\textbf {\bibinfo {volume} {16}},\
  \bibinfo {pages} {103011} (\bibinfo {year} {2014})}\BibitemShut {NoStop}%
\bibitem [{\citenamefont {Mohammady}\ \emph {et~al.}(2016)\citenamefont
  {Mohammady}, \citenamefont {Mohseni},\ and\ \citenamefont
  {Omar}}]{mohammady2016minimising}%
  \BibitemOpen
  \bibfield  {author} {\bibinfo {author} {\bibfnamefont {M.~H.}\ \bibnamefont
  {Mohammady}}, \bibinfo {author} {\bibfnamefont {M.}~\bibnamefont {Mohseni}},\
  and\ \bibinfo {author} {\bibfnamefont {Y.}~\bibnamefont {Omar}},\ }\bibfield
  {title} {\bibinfo {title} {Minimising the heat dissipation of quantum
  information erasure},\ }\href {https://doi.org/10.1088/1367-2630/18/1/015011}
  {\bibfield  {journal} {\bibinfo  {journal} {New J. Phys.}\ }\textbf {\bibinfo
  {volume} {18}},\ \bibinfo {pages} {015011} (\bibinfo {year}
  {2016})}\BibitemShut {NoStop}%
\bibitem [{\citenamefont {Mart{\'\i}nez}\ \emph {et~al.}(2016)\citenamefont
  {Mart{\'\i}nez}, \citenamefont {Petrosyan}, \citenamefont {Gu{\'e}ry-Odelin},
  \citenamefont {Trizac},\ and\ \citenamefont
  {Ciliberto}}]{martinez2016engineered}%
  \BibitemOpen
  \bibfield  {author} {\bibinfo {author} {\bibfnamefont {I.~A.}\ \bibnamefont
  {Mart{\'\i}nez}}, \bibinfo {author} {\bibfnamefont {A.}~\bibnamefont
  {Petrosyan}}, \bibinfo {author} {\bibfnamefont {D.}~\bibnamefont
  {Gu{\'e}ry-Odelin}}, \bibinfo {author} {\bibfnamefont {E.}~\bibnamefont
  {Trizac}},\ and\ \bibinfo {author} {\bibfnamefont {S.}~\bibnamefont
  {Ciliberto}},\ }\bibfield  {title} {\bibinfo {title} {Engineered swift
  equilibration of a {B}rownian particle},\ }\href
  {https://doi.org/10.1038/NPHYS3758} {\bibfield  {journal} {\bibinfo
  {journal} {Nat. Phys.}\ }\textbf {\bibinfo {volume} {12}},\ \bibinfo {pages}
  {843} (\bibinfo {year} {2016})}\BibitemShut {NoStop}%
\bibitem [{\citenamefont {Chupeau}\ \emph {et~al.}(2018)\citenamefont
  {Chupeau}, \citenamefont {Besga}, \citenamefont {Gu{\'e}ry-Odelin},
  \citenamefont {Trizac}, \citenamefont {Petrosyan},\ and\ \citenamefont
  {Ciliberto}}]{chupeau2018thermal}%
  \BibitemOpen
  \bibfield  {author} {\bibinfo {author} {\bibfnamefont {M.}~\bibnamefont
  {Chupeau}}, \bibinfo {author} {\bibfnamefont {B.}~\bibnamefont {Besga}},
  \bibinfo {author} {\bibfnamefont {D.}~\bibnamefont {Gu{\'e}ry-Odelin}},
  \bibinfo {author} {\bibfnamefont {E.}~\bibnamefont {Trizac}}, \bibinfo
  {author} {\bibfnamefont {A.}~\bibnamefont {Petrosyan}},\ and\ \bibinfo
  {author} {\bibfnamefont {S.}~\bibnamefont {Ciliberto}},\ }\bibfield  {title}
  {\bibinfo {title} {Thermal bath engineering for swift equilibration},\ }\href
  {https://doi.org/10.1103/PhysRevE.98.010104} {\bibfield  {journal} {\bibinfo
  {journal} {Phys. Rev. E}\ }\textbf {\bibinfo {volume} {98}},\ \bibinfo
  {pages} {010104} (\bibinfo {year} {2018})}\BibitemShut {NoStop}%
\bibitem [{\citenamefont {Gingrich}\ \emph {et~al.}(2016)\citenamefont
  {Gingrich}, \citenamefont {Rotskoff}, \citenamefont {Crooks},\ and\
  \citenamefont {Geissler}}]{gingrich2016near}%
  \BibitemOpen
  \bibfield  {author} {\bibinfo {author} {\bibfnamefont {T.~R.}\ \bibnamefont
  {Gingrich}}, \bibinfo {author} {\bibfnamefont {G.~M.}\ \bibnamefont
  {Rotskoff}}, \bibinfo {author} {\bibfnamefont {G.~E.}\ \bibnamefont
  {Crooks}},\ and\ \bibinfo {author} {\bibfnamefont {P.~L.}\ \bibnamefont
  {Geissler}},\ }\bibfield  {title} {\bibinfo {title} {Near-optimal protocols
  in complex nonequilibrium transformations},\ }\href
  {https://doi.org/10.1073/pnas.1606273113} {\bibfield  {journal} {\bibinfo
  {journal} {Proc. Natl. Acad. Sci. U.S.A.}\ }\textbf {\bibinfo {volume}
  {113}},\ \bibinfo {pages} {10263} (\bibinfo {year} {2016})}\BibitemShut
  {NoStop}%
\end{thebibliography}
\end{document}


\title{Supplementary Material for Finite-time Landauer Principle}

\author{Karel Proesmans}
\affiliation{Department of Physics, Simon Fraser University, Burnaby, B.C., V5A 1S6, Canada}
\affiliation{Hasselt University, B-3590 Diepenbeek, Belgium}
\author{Jannik Ehrich}
\affiliation{Department of Physics, Simon Fraser University, Burnaby, B.C., V5A 1S6, Canada}
\author{John Bechhoefer}
\affiliation{Department of Physics, Simon Fraser University, Burnaby, B.C., V5A 1S6, Canada}

\date{\today}
\maketitle
In this Supplement, we give details for comparing our results on the minimum average amount of work needed for erasure with previous numerical and experimental studies.  Those studies were typically conducted in the slow-driving limit, where $W_\textrm{min} \approx W_\textrm{min,leq}$.  In physical units, these results all take the form
\begin{equation}
    \frac{W_\textrm{min}}{k_\textrm{B}T} - \ln 2 = a \frac{\textrm{Var}(x)}{D\tau} \,,
\label{eq:workasymptotic}
\end{equation}
where $a$ is a dimensionless constant whose value measures the cost of a given protocol.  

In experiments and simulations, the measured average work $W$ has generally been found to follow the form given in Eq.~\eqref{eq:workasymptotic}.  Although the focus of previous work has been to evaluate the $\tau \to \infty$ limit, to see whether the asymptotic work is consistent with the Landauer bound of $\ln 2$, here we focus on the measured value of $a$, comparing it to a predicted minimum value, $a_\textrm{min}$.  The results described here are collected in Table 1 in the paper.

We consider five different cases.  Three are experiments where a single colloidal particle is immersed in water.  The other two are theoretical studies supported by simulations.  We begin with the three experiments, which all use a potential at least approximately described by the double-well form explored in the main text,
\begin{equation}
    V_0(x) = E_\textrm{b}
        \left[ \left( \frac{x}{x_\textrm{m}} \right)^2 - 1 \right]^2 \,.
\label{Vdw}
\end{equation}

\begin{itemize}[leftmargin=*]
\item \textbf{Bérut et al.} \cite{Berut2012experimental,Berut2013detailed,Berut2015information}: The potential $V_0(x)$, defined by an optical tweezer that rapidly hops back and forth between two positions ($\pm x_\textrm{m}$).  The distance between the wells is given by $x_\textrm{m}=0.45$ \textmu m, and the height of the energy barrier is $E_\textrm{b} = 10 \, k_\textrm{B}T$. Furthermore, they report $D=0.25$ \textmu m$^2$ s$^{-1}$. This leads to $\tau_0 = \textrm{Var}(x)/D=0.8$ s.  On the other hand, they report
\begin{equation}
    \frac{W}{k_\textrm{B}T}-\ln 2\approx \frac{8.15 \, \textrm{s}}{\tau}=10.3\frac{\textrm{Var}(x)}{D\tau} \,.
\end{equation}
Meanwhile, one can numerically evaluate $W_{\textrm{min,leq}}$ from Eq.~(3) from the main text (see also Ref.~\cite{proesmans20}),
\begin{equation}
    \frac{W_{\textrm{min}}}{k_\textrm{B}T}-\ln 2\approx\frac{W_{\textrm{min,leq}}}{k_\textrm{B}T}-\ln 2=1.80\frac{\textrm{Var}(x)}{D\tau} \,.
\end{equation}
    
\item \textbf{Gavrilov et al.}~\cite{gavrilov2017direct}: The potential $V_0(x)$ again has the form of Eq.~\eqref{Vdw}, with $x_\textrm{m}=0.77$ \textmu m, $D=0.23$ \textmu m$^2$ s$^{-1}$ and $E_\textrm{b} = 13 \, k_\textrm{B}T$.  They report 
\begin{equation}
    \frac{W}{k_\textrm{B}T}-\ln 2 
        \approx \frac{1.765(2x_\textrm{m})^2}{Dt}
        \approx 7.20\frac{\textrm{Var}(x)}{D\tau} \,,
\end{equation}
while the optimal protocol again gives
\begin{equation}
    \frac{W_{\textrm{min}}}{k_\textrm{B}T}-\ln 2=1.82\frac{\textrm{Var}(x)}{Dt} \,.
\end{equation}

\item \textbf{Jun et al.} \cite{jun2014high}:  The potential $V_0(x)$ was imposed by a feedback trap, with $x_\textrm{m}=2.5$ \textmu m,  $E_\textrm{b} = 13 \, k_\textrm{B}T$ and $D=1.7$ \textmu m$^2$ s$^{-1}$. From the experiment,
\begin{equation}
    \frac{W}{k_\textrm{B}T} - \ln 2 
        \approx \frac{1.39(2x_\textrm{m})^2}{Dt} 
    \approx 5.67\frac{\textrm{Var}(x)}{D\tau} \,.
\end{equation}
On the other hand, an exact calculation leads to
\begin{equation}
    \frac{W_{\textrm{min}}}{k_\textrm{B}T}- \ln2 
        \approx 1.82\frac{\textrm{Var}(x)}{Dt} \,.
\label{wopber}
\end{equation}
    
\end{itemize}

We conclude with two theoretical studies:
\begin{itemize}[leftmargin=*]
\item \textbf{Zulkowski \& DeWeese} \cite{zulkowski2014optimal}:  The potential $V_0(x)$ has two flat wells of length $\ell$, separated by a thin wall, as shown in Fig.~\ref{fig:flatwell}.  One can easily check that
\begin{equation}
    \textrm{Var}(x)=\frac{1}{2\ell}
        \int^\ell_{-\ell} \dd{x} \, x^2 = \frac{\ell^2}{3} \,.
\end{equation}

\begin{figure}[ht]
    \includegraphics[width=3.0in]{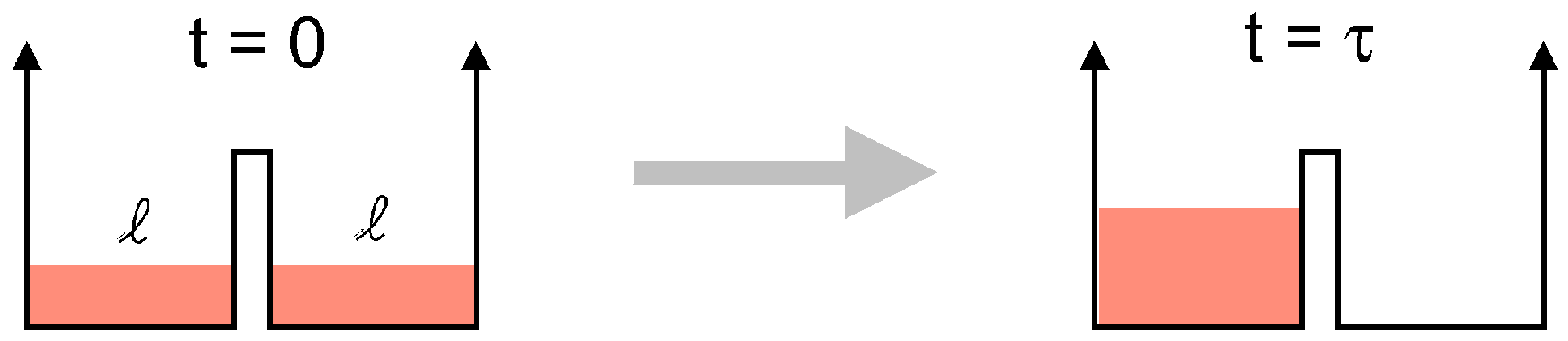}
    \caption{Bit erasure in a flat-well potential, with $\tau \gg 1$. Pink shaded areas represent $p_0(x)$ at left and $p_\tau(x)$ at right.}
\label{fig:flatwell}
\end{figure}

The authors showed that, if one has only control over the energy levels of the separate wells, the minimum amount of work associated with the full erasure of a bit is given by
\begin{equation}
    \frac{W}{k_\textrm{B}T}-\ln 2 
        = \frac{3\, \textrm{Var}(x)}{D\tau}\left[ \left(\sqrt{2}-1\right)^2+1\right] \,.
\label{wopZD}
\end{equation}

On the other hand, erasure under full control can also be obtained using our framework. Indeed, as illustrated in Fig.~\ref{fig:flatwell}, we have $p_0(x)=1/(2\ell)$ and
\begin{equation}
    p_\tau(x)=
        \begin{cases}
            \frac{1}{\ell} &x<0 \\
            0 & x>0 \,,
         \end{cases}
\end{equation}
as the system ends up in local equilibrium in the slow-driving limit.  A straightforward calculation, using Eq.~(3) from the main text (see also Ref.~\cite{proesmans20}) leads to
\begin{equation}
     \frac{W_{\textrm{min}}}{k_\textrm{B}T}-\ln 2 
        = \frac{\textrm{Var}(x)}{D\tau} \,.
\label{fwopt}
\end{equation}

\item \textbf{Boyd et al.} \cite{boyd2018shortcuts}: The probability distribution is always kept in local equilibrium,
\begin{equation}
    p(x,t)=
        \begin{cases}
            [1-b(t)]p_0(x) & x<0 \\
            b(t)p_0(x) & x>0 \,,
        \end{cases}
\end{equation}
where $b(t)=p(x>0,t)$ is the probability for the bit to be in state $1$ at time $t$. The authors show that the required work to erase a bit under such a constraint is given by
\begin{equation}
    \frac{W}{k_\textrm{B}T}-\ln 2-\epsilon\ln \epsilon 
        - (1-\epsilon) \ln (1-\epsilon)
    =f_1 \left[ p_0(x) \right] f_2 \left[ b(t) \right] \,,
\end{equation}
with \cite{note1}
\begin{align}
    f_1\left[p_0(x)\right] &= \frac{2}{D}
        \int^{\infty}_0 \dd{x}\, \int^x_0 \dd{x'} \, 
        \int^{-x'}_{-\infty} \dd{x''} \, \frac{p(x)p(x'')}{p(x')} 
        \nonumber \\
    f_2\left[b(t)\right]&= \int_0^{\tau}\dd{t}\, 
        \frac{[\partial_tb(t)]^2}{b(t)-b(t)^2} \,.
\end{align}
For the system discussed by Zulkowski and DeWeese \cite{zulkowski2014optimal}, the initial density is $p_0(x)=1/(2\ell)$, and one can easily check that
\begin{equation}
    f_1\left[p_0(x)\right]=\frac{\textrm{Var}(x)}{D} \,.
\end{equation}
On the other hand, following \cite{boyd2018shortcuts}, we can set
\begin{equation}
    b(t)=\frac{1}{2}\cos^2\left(\frac{\pi t}{2\tau}\right) \,,
\end{equation}
which satisfies $b(0)=1/2$ and $b(\tau)=0$.  This leads to
\begin{equation}
    f_2\left[b(t)\right]=\frac{\pi^2}{4\tau}
        \left[\left(\sqrt{2}-1\right)^2+1\right] \,.
\label{f2b}
\end{equation}
Using these expressions for $f_1$ and $f_2$, we find
\begin{equation}
    \frac{W}{k_\textrm{B}T}-\ln2 =       
        \frac{\pi^2(2-\sqrt{2})}{2} \frac{\textrm{Var}(x)}{D\tau}
    \approx 2.89 \frac{\textrm{Var}(x)}{D\tau} \,.
\end{equation}
    
We can also apply the protocol of Boyd \emph{et al}.~to a bit in a potential of the form of Eq.~\eqref{Vdw}.  We choose parameters $E_\textrm{b} = 10 \, k_\textrm{B}T$, $x_\textrm{m}=0.45$ \textmu m, $D=0.25$ \textmu m$^2$ s$^{-1}$ and $\epsilon=0$, to match those of Bérut \emph{et al}. \cite{Berut2012experimental}. With these numbers, we find
\begin{equation}
    f_1\left[p_0(x)\right]\approx 1312 \frac{\textrm{Var}(x)}{D} \,,
\end{equation}
and $f_2\left[b(t)\right]$ is the same as in Eq.~\eqref{f2b}.  We then conclude that 
\begin{equation}
     \frac{W}{k_\textrm{B}T}-\ln 2=3786\frac{\textrm{Var}(x)}{D\tau} \,,
\end{equation}
which exceeds the optimal value, Eq.~\eqref{wopber}, by a factor $\approx 2000$. Similar results hold for the parameters corresponding to the experiments of Yun \emph{et al}.~\cite{jun2014high} and Gavrilov \emph{et al}.~\cite{gavrilov2017direct}.
\end{itemize}